\documentclass{aa}  
\bibpunct{(}{)}{;}{a}{}{,} 
\usepackage{natbib}
\usepackage{bigstrut}
\usepackage{graphicx}
\usepackage{txfonts}
\begin{document}

   \title{Oxygen isotopic ratios in intermediate-mass red giants}

   \author{T. Lebzelter\inst{1}\thanks{Visiting astronomer, Kitt Peak National Observatory, National Optical Astronomy Observatory}
          \and
          O. Straniero\inst{2}
          \and
          K.H. Hinkle\inst{3}
          \and
          W. Nowotny\inst{1}
          \and
          B. Aringer\inst{1,4}
          }

   \institute{Department of Astrophysics, University of Vienna,
              T\"urkenschanzstrasse 17, A-1180 Vienna\\
              \email{thomas.lebzelter@univie.ac.at}
         \and
             INAF, Osservatorio Astronomico di Collurania, 64100 Teramo, Italy\\
             \email{straniero@oa-teramo.inaf.it}
         \and
             National Optical Astronomy Observatory, P.O. Box 26732, Tucson, AZ 85726, USA \\
             \email{khinkle@noao.edu}
         \and
             Dipartimento di Fisica e Astronomia Galileo Galilei, Universita di Padova, Vicolo dell'Osservatorio 3, I-35122 Padova, Italy\\
             }

   \date{Received ; accepted }

 
  \abstract
   {The abundances of the three main isotopes of oxygen are altered in the course of the CNO-cycle. When the first dredge-up
   mixes the burning products to the surface, the nucleosynthesis processes can be probed by measuring oxygen isotopic 
   ratios.}
   {By measuring $^{16}$O/$^{17}$O and $^{16}$O/$^{18}$O in red giants of known mass we compare the isotope ratios with
   predictions from stellar and galactic evolution modelling.}
   {Oxygen isotopic ratios were derived from the $K$-band spectra of six red giants.  The sample red giants are open cluster members with known 
masses of between 1.8 and 4.5\,M$_{\sun}$.
The abundance determination employs synthetic spectra calculated with
   the COMARCS code. The effect of uncertainties in the nuclear reaction rates, the mixing length, and of a change in
   the initial abundance of the oxygen isotopes was determined by a set of nucleosynthesis and mixing models using the
   FUNS code.}
   {The observed $^{16}$O/$^{17}$O ratios are in good agreement with the model results, even if the measured values do not 
   present clear evidence of a variation with the stellar mass.  The observed
   $^{16}$O/$^{18}$O ratios are clearly lower than the predictions from our reference model. Variations in
   nuclear reaction rates and mixing length parameter both have only a very weak effect on the predicted values.
   The $^{12}$C/$^{13}$C ratios of the K giants studied implies the absence of extra-mixing in these objects.}
   {A comparison with galactic chemical evolution models indicates that the $^{16}$O/$^{18}$O abundance ratio underwent a faster decrease 
than predicted.  To explain the observed ratios, the most likely scenario is a higher initial $^{18}$O abundance combined with 
   a lower initial $^{16}$O abundance.  Comparing the measured $^{18}$O/$^{17}$O ratio with the corresponding value for the ISM
   points towards an initial enhancement of $^{17}$O as well.
   Limitations imposed by the observations prevent this from being a conclusive result.}


   \maketitle
%

\section{Introduction}
The ratios of the abundances of the three stable isotopes of oxygen,
$^{16}$O/$^{17}$O and $^{16}$O/$^{18}$O, are important indicators
of the nucleosynthesis and mixing in the interiors of stars
\citep{1992PhR...210..367D}. In an H-burning environment
$^{16}$O(p,$\gamma$)$^{17}$F($\beta^{+}$)$^{17}$O causes a steep
increase in the $^{17}$O  abundance until equilibrium is reached
by the competitive $^{17}$O(p,$\alpha$)$^{14}$N process
\citep{1990A&A...240...85L}. Owing to the steep abundance gradient
of $^{17}$O resulting from incomplete CNO-cycle burning, the surface
abundance of the $^{17}$O isotope is highly sensitive to the precise
depth of convection and the mixing profile in red giants
\citep{1999ApJ...510..232B}. $^{18}$O is primarily produced via
$^{14}$N($\alpha$,$\gamma$)$^{18}$F($\beta^{+}$)$^{18}$O
\citep[e.g.][]{2003hic..book.....C}.  This reaction takes place in
the He-burning shells of massive stars. The isotope is destroyed
in hydrogen-burning stars by $^{18}$O(p,$\alpha$)$^{15}$N and destroyed
during He burning by $^{18}$O($\alpha$,$\gamma$)$^{22}$Ne.
In low and intermediate mass stars, the $^{18}$O abundance thus
reflects the initial abundance of this isotope and its destruction
rate during hydrogen burning.  The reaction rates for the processes
including the various oxygen isotopes are relevant for later steps
in the stellar nucleosynthesis, in particular the s-process \citep[see
the discussion in, e.g.,][]{2014ApJ...785...77S}.

The solar values for $^{16}$O/$^{17}$O and $^{16}$O/$^{18}$O are
2700 and 498 \citep{2009LanB...4B...44L}, respectively.  Based on
the interplay between the production, mixing, and destruction of the
oxygen isotopes, stellar evolution models predict a correlation
between oxygen isotopic ratios and stellar mass for red giants after
the first dredge-up
\citep{1994ApJ...430L..77B,1994A&A...285..915E,2003MNRAS.340..763S,2014PASA...31...30K}.
The various models agree on a steep decrease in the ratio
$^{16}$O/$^{17}$O between 1 and around 2\,$M_{\sun}$ followed by a
weak increase in this ratio for higher masses. At the same time,
the $^{16}$O/$^{18}$O value after the first dredge-up is expected
to show a moderate increase for stars below 2\,$M_{\sun}$ and a
constant value afterwards.  There are small quantitative differences
for the predicted $^{16}$O/$^{18}$O ratio between the models. A
direct comparison of abundances is hampered by variations in the
starting values chosen by the various models.

For testing the predictions, RGB stars of known mass are crucial.
\citet{1994A&A...285..915E} attempted to do such a comparison with
the help of a set of bright field stars. However, as noted by the
author of that study, masses are rather uncertain for these objects.
Accordingly, a convincing observational test of the models was not
possible.

The CO vibration-rotation 4.6 $\mu$m fundamental or 2.3 $\mu$m first
overtone bands were used in early determinations of O isotopes in
red giants.  The early determinations were limited to bright stars
due to instrumental constraints \citep[see][and references
therein]{1988ApJ...325..768H,1990ApJS...72..387S}.  Contemporary
instrumentation is much more sensitive, allowing the study of weaker
targets at high spectral resolution and high S/N.  It is now possible
to observe open-cluster (OC) red giants where stellar mass can be
determined from the cluster age.  Exploring the causes for oxygen
isotopic ratios that deviate significantly from predictions based
on the solar value provides insight into the cosmic matter cycle.

\section{Sample definition and observations}

For the target selection, we chose the list of red giants in galactic
OC provided by \citet{1989ApJ...347..835G} plus two stars in NGC
7789 from \citet{2011A&A...531A.165P}.  The total sample selected
on the basis of observability (see below) and stellar mass includes
seven stars.  \citet{1989ApJ...347..835G} provides the masses of
the stars on the giant branch, and we used these values for our
analysis. For NGC 7789 there are two recent age determinations of
1.4\,Gyr \citep{2008ApJ...676..594K} and 1.8\,Gyr
\citep{2011AJ....142...59J}. When using isochrones from
\citet{2012MNRAS.427..127B}, the turn-off mass is 1.8$\pm$0.2\,$M_{\sun}$.
The metallicities of the selected clusters are close to solar
\citep{1989ApJ...347..835G,2011A&A...531A.165P}. Our sample of red
giants is listed in Table\,\ref{t:sample}.

Subsequent to taking the observations, we found that HD\,49050 is
no longer considered a member of NGC\,2287 \citep{2008A&A...485..303M}.
On the other hand, cluster membership of HD\,16068 in Tr 2 and
HD\,27292 in NGC\,1545 was confirmed by \citet{2011MNRAS.417..649Z}.
For HD\,68879, \citet{2008AJ....136..118F} list a cluster membership
probability of more than 96\,\%.  For the other stars, their radial
velocities are very close to the respective mean cluster velocity
\citep{2008A&A...485..303M}, which supports their membership.  A
re-examination of the kinematics of the UMa group based on Hipparcos
data leaves the membership of HD\,30834 unclear
\citep{2003AJ....125.1980K}.

We obtained several small pieces of high-resolution spectra in the
$H$ and $K$ bands using the Phoenix spectrograph
\citep{2000SPIE.4008..720H} at the Kitt Peak 2.1m telescope.
Observations were obtained in December 2013 and January 2014. The
standard infrared observing procedure with two nodding positions
was applied. Telluric lines were removed by ratioing the spectra
to spectra of hot stars observed at similar airmass. The telluric
lines in the hot star spectra were also used for the wavelength
calibration.

\section{Isotopic ratios} \label{isoratio}

Our goal was to derive the $^{16}$O/$^{17}$O, $^{16}$O/$^{18}$O,
and $^{12}$C/$^{13}$C ratios for our seven target stars.  Owing to
the very limited spectral range of about 100\,{\AA} covered by a
single observation, the number of spectral features available for
our analysis is smaller than in the earlier studies based on FTS
scans of the whole $K$-band region. To measure the $^{17}$O abundance
we used two to four 2-0 band C$^{17}$O lines located near 4285.4
cm$^{-1}$. These lines are largely unblended and can
be easily identified for a stellar temperature of up to 4200\,K.

Determining the $^{18}$O abundance from $K$-band spectra is more
difficult. We ultimately used a region around 4226\,cm$^{-1}$, which
was used by \citet{2010ApJ...714..144G} for their study of oxygen
isotopes in R\,CrB stars. The most usable line of C$^{18}$O in this
region is the 2-0~R23 line, which is not affected significantly by
telluric lines. However, this line is affected by an unidentified
blend, so its usability for abundance determination is limited to $T_{\rm
eff}$<4100\,K and log\,$g$<2.0.  An attempt to use the
2-0 band head of C$^{18}$O, as in \citet{1990ApJS...72..387S}, for instance,
failed since the band head is located too close to the transmission
cut-off of the Phoenix order separation filter. 

Complimentary information on mixing in the stellar atmosphere was
derived by measuring the $^{12}$C/$^{13}$C ratio for our sample
stars.  Several first overtone $^{12}$C$^{16}$O and $^{13}$C$^{16}$O
lines near 4246\,cm$^{-1}$ and 4262\,cm$^{-1}$ were used.  We also
attempted observations of the CO fundamental spectrum near 4.6\,$\mu$m
where many CO lines from all three oxygen isotopes can be found.
However, lower flux levels, line blending, and difficulties in
setting the continuum level limited the use of these data. The data
presented in this paper are part of a larger observing programme
on oxygen abundances in cool red giants. Details will be presented
elsewhere.

Isotopic ratios were computed using spectrum synthesis techniques.
For the stellar atmospheric structure, we used the hydrostatic spherical
COMARCS models described in \citet{2009A&A...503..913A}.  Synthetic
spectra were calculated with the COMA code using the same set of
opacity data. Calculations were done under the assumption of LTE.
We adopted values for solar composition provided by
\citet{2009MmSAI..80..643C}.  Abundances were determined by a direct
comparison between observed and synthetic spectra using both visual
inspection and the measured line depths. For a more detailed
description of our approach, we refer to earlier applications to the
fitting of similar high resolution spectra given in
\citet{2008A&A...486..511L} and \citet{2009A&A...502..913L}. 
Line positions throughout the studied spectral range have been
improved by using the work of \citet{1995iaas.book.....H} and 
\citet{2010JQSRT.111..521H}. Data for the lines used in this 
analysis are listed in Table~\ref{t:linelist}.

\begin{figure}
\resizebox{\hsize}{!}{\includegraphics[angle=90]{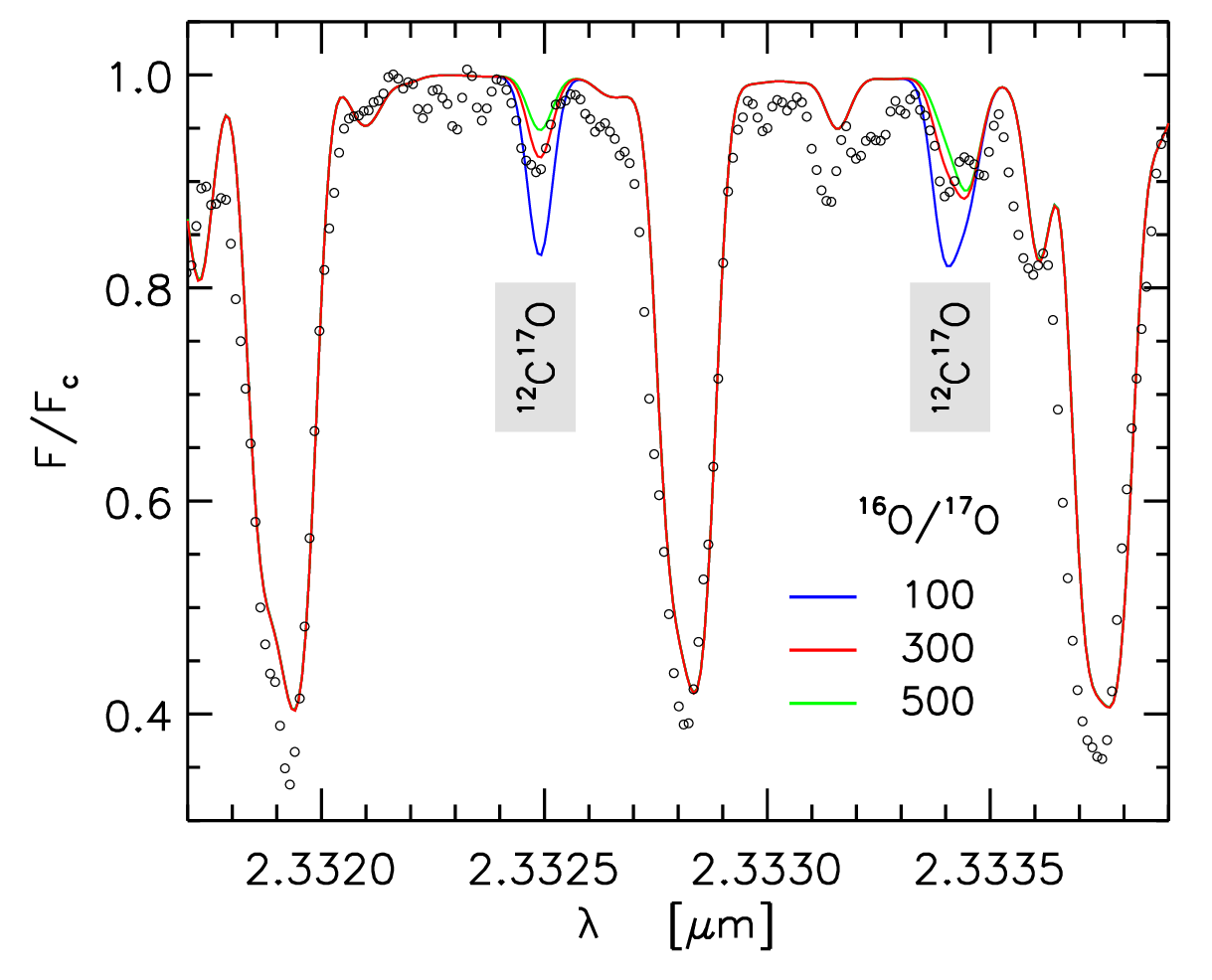}}
\caption{Observed spectrum of HD 27292 (dots) showing two $^{12}$C$^{17}$O lines. The three
synthetic spectra (solid lines) are for $T_{\rm eff}$=4000\,K and log$g$=1.5 with $^{16}$O/$^{17}$O=100, 300, and 500.}
\label{f:17ospec}
\end{figure}

\begin{figure}
\resizebox{\hsize}{!}{\includegraphics[angle=90]{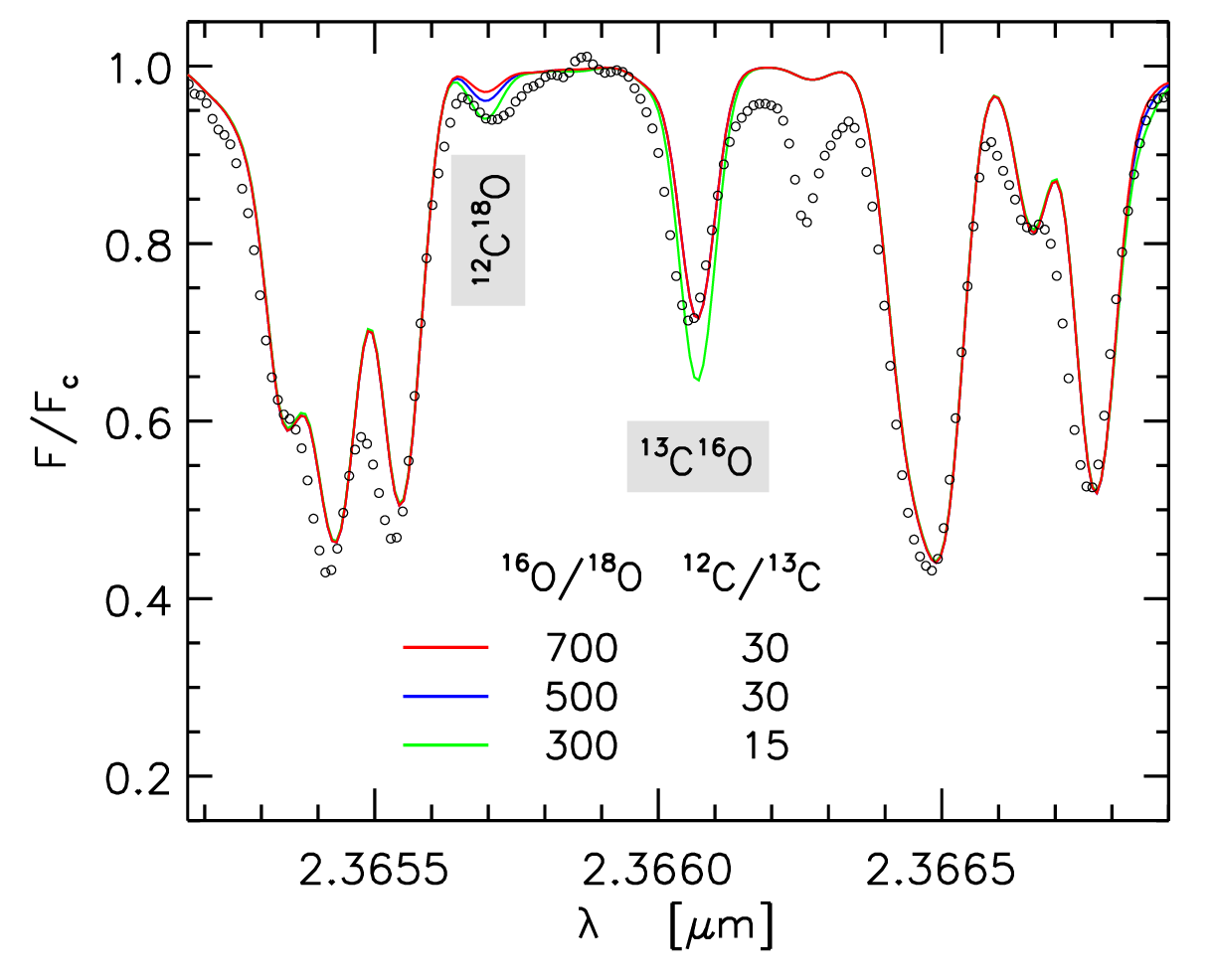}}
\caption{Observed spectrum of HD 27292 (dots) with a $^{12}$C$^{18}$O line and a $^{13}$C$^{16}$O line marked. The three
synthetic spectra (solid lines) are for $T_{\rm eff}$=4000\,K and log$g$=1.5 with $^{16}$O/$^{18}$O=300, 500, and 700 and 
$^{12}$C/$^{13}$C=15 and 30, respectively.}
\label{f:18ospec}
\end{figure}

The line strengths depend on the stellar temperature, surface
gravity, chemical composition, and abundance of each isotope.  For
the stars taken from the list of \citet{1989ApJ...347..835G} ,we used
the values for $T_{\rm eff}$ and log\,$g$ given there.
\citet{2012MNRAS.427..343M} independently determined  $T_{\rm eff}$ values
for HD\,27292 (3844\,K), HD\,30834 (4247\,K), and HD 68879 (4552\,K),
which are in reasonable agreement with the values given by
Gilroy\footnote{\citet{2013A&A...549A.129C} list a rather high
temperature of 6427\,K for HD\,16068. They refer to a parameter
determination by \citet{2008yCat.5128....0H}. However, we could not
find this star in the latter catalogue. Therefore, we suppose there
is a mistake in \citet{2013A&A...549A.129C}.}.  For the two stars
in NGC\,7789, the stellar parameters were taken from the study of
\citet{2011A&A...531A.165P}.  Since all clusters have a metallicity
close to solar
\citep{1989ApJ...347..835G,2011AJ....142...59J,2011MNRAS.414.1227M}, we
calculated our grid of synthetic spectra only for solar composition.
For C/O we chose a fixed value of 0.3 to resemble a typical value
for post first dredge-up composition. The microturbulence $\xi$
was set 2.0\,km\,s$^{-1}$, which is a typical value for these
stars according to \citet{1989ApJ...347..835G}. The macro turbulence
was set to values between 1.5 and 2.5\,kms$^{-1}$ to optimize the
fit of all stellar lines within the observed wavelength range. All parameters taken
from the literature were cross-checked and confirmed with our
spectra. The finally chosen values for $T_{\rm eff}$ and log\,$g$
are given in columns 4 and 5 of Table\,\ref{t:sample}.  Examples
for the spectral fit are shown in Figs.~\ref{f:17ospec} and
\ref{f:18ospec}. We note that our synthetic spectra cannot fit the cores
of the strongest CO lines (lowest excitation) properly (e.g.~Fig.\ref{f:17ospec}). 
This difficulty has also been encountered by
other authors \citep[e.g.][]{2008A&A...489.1271T} and likely results from limitations
of hydrostatic model atmospheres in describing the outer layers of a red giant
properly.

The resulting oxygen isotopic ratios are listed in Table\,\ref{t:sample}.
The two stars in the cluster NGC\,7789 show similar values. For the
hottest star in our sample, HD\,68879, we could not determine
$^{16}$O/$^{17}$O and $^{16}$O/$^{18}$O because the CO lines were
too weak.  For HD\,30834 we could not measure $^{16}$O/$^{18}$O because the
contribution of the C$^{18}$O line to the blend could not be
constrained properly.  For this star, the value given for $^{16}$O/$^{18}$O
in Table\,\ref{t:sample} is a lower limit. We also determined carbon
isotopic ratios for all our stars. The carbon isotope results are
in good agreement with the values determined by
\citet{1989ApJ...347..835G} (see Table\,\ref{t:sample}).

To interpret the $^{16}$O/$^{17}$O and $^{16}$O/$^{18}$O ratios, we
need to check whether a solar $^{16}$O abundance value is appropriate
for our sample stars. We did this by computing model spectra with
altered oxygen abundances and with stellar parameters and carbon
isotopic ratio as derived in the previous step. Since C/O\,<\,1 the
strengths of the CO lines are dependent on the carbon abundance and
the isotopic ratios of oxygen but not on the total oxygen abundance.
To derive the oxygen abundance, OH vibration-rotation lines in the
$H$ band were modelled. The $H$-band spectra around 6072\,cm$^{-1}$
were obtained for six of the K giants in Table\,\ref{t:sample}.  We
modelled the three least blended $^{16}$OH lines, 4-2\,P$_{\rm
1f}$\,6.5, 2-0\,P$_{\rm 1f}$\,16.5, and 2-0\,P$_{\rm 2e}$\,15.5.
The resulting abundances listed in Table\,\ref{t:sample} were all
sub-solar by -0.16 to -0.36\,dex.  The uncertainties are dominated
by the temperature sensitivity of the chosen OH lines.  As noted
above, the total oxygen abundance has no effect on the strengths
of the CO lines.  We also do not expect that the oxygen abundance
will affect the atmospheric structure because photospheric H$_{2}$O is not present in the temperature
range of our sample stars.

According to \citet{1989ApJ...347..835G}, the uncertainties of the
stellar parameters are $\Delta\,T_{\rm eff}$~$\pm$150\,K,
$\Delta$\,log\,$g$~$\pm$0.3, $\Delta\,\xi$~$\pm$0.2\,kms$^{-1}$,
and $\Delta$\,[Fe/H]~$\pm$0.2. Alternative parameter values found
in the literature agree within these error bars. For each star we derived
the isotopic ratios by changing the stellar
parameters within these ranges.  Some combinations, however, were
not included when estimating the uncertainties because the corresponding
synthetic spectra do not provide a good fit to the observations.
Since the continuum level in the $K$ band is well defined 
for these stars, we did not consider it to have an impact on the
error budget. The final uncertainties, given in Table\,\ref{t:sample},
were then determined by combining the maximum differences from changes
in the stellar parameters with the scatter resulting when various
lines were used for the abundance determination. The latter was
typically a factor of 2 smaller than the uncertainties from the
stellar parameters.  As noted above we did not determine
the C/O ratios from our spectra but set them to a fixed value of 0.3 in our analysis.
We explored the effect of changing the C/O ratio by $\pm$0.1 dex on
the derived oxygen isotopic ratios. Such a change somewhat modifies the strength of
all CO lines, but primarily those of weak or moderate strength. To achieve a good
fit of the whole spectrum again, the stellar parameters have to be adapted so that
the net effect on the isotopic ratios is small and covered by the error budget 
derived above. The effect of changing the macroturbulence value by 1\,kms$^{-1}$
is a few percent on the resulting isotopic ratio.

For HD\,16068 and HD\,27292, we also observed the $M$-band spectrum
near 2154\,cm$^{-1}$. As mentioned above, the $M$-band spectra could
not be used to derive reliable isotopic ratios. However, attempting
to achieve a reasonable fit to the 4.6\,$\mu$m spectra limits the
$^{16}$O/$^{17}$O and $^{16}$O/$^{18}$O ratios to a range between
200 and 600, in agreement with the findings from the $K$ band.

\begin{table*}
\caption{Stellar parameters and isotopic ratios. Data in columns 3 to 6 are taken from the literature.} 
\label{t:sample} 
\centering 
\begin{tabular}{l l c c c c c c c c c} 
\hline\hline 
Cluster & Star & Mass [$M_{\sun}$] & $T_{\rm eff}$ [K] & log\,$g$ & $^{12}$C/$^{13}$C Lit. & Ref. & [$^{16}$O/H] & $^{16}$O/$^{17}$O & $^{16}$O/$^{18}$O & $^{12}$C/$^{13}$C \\
\hline
NGC\,7789 & 971 & 1.8$\pm$0.2 & 3700 & 1.2 & -- & (2) & -0.18$\pm$0.1 & 300$_{-70}^{+100}$ & 340$_{-50}^{+150}$ & 15$\pm$8 \bigstrut\\
         & 415 & 1.8$\pm$0.2 & 3800 & 1.2 & -- & (2) & -0.16$\pm$0.1 & 250$_{-70}^{+100}$ & 270$_{-40}^{+150}$ & 19$\pm$8 \bigstrut\\
NGC\,2548 & \object{HD 68879} & 2.6$\pm$0.3 & 4600 & 2.2 & 25 & (1) & -- & -- & -- & 16$\pm$7\bigstrut\\
NGC\,1545 & \object{HD 27292} & 2.8$\pm$0.2 & 4000 & 1.5 & 25 & (1) & $-$0.33$\pm$0.1 & 240$_{-50}^{+90}$ & 340$_{-70}^{+110}$ & 27$\pm$10\bigstrut\\
UMa Grp. & \object{HD 30834} & 2.9$\pm$0.4 & 4000 & 2.2 & 26 & (1) & $-$0.36$\pm$0.1 & 260$_{-50}^{+90}$ & $>$200 & 22$\pm$7\bigstrut\\ 
NGC\,2287 & \object{HD 49050} & --\tablefootmark{a} & 4100 & 1.6 & 23 & (1) & $-$0.19$\pm$0.1 & 300$_{-50}^{+150}$ & --\tablefootmark{b} & 30:\bigstrut\\
Tr\,2 & \object{HD 16068} & 4.5$\pm$0.4 & 4000 & 1.5 & 30 & (1) & $-$0.28$\pm$0.1 & 280$_{-50}^{+90}$ & 370$_{-90}^{+100}$ & 28$\pm$10\bigstrut\\
\hline
\end{tabular}
\tablebib{
(1) \citet{1989ApJ...347..835G}; (2) \citet{2011A&A...531A.165P}.
}
\tablefoot{
\tablefoottext{a}{Star is probably not a cluster member.}
\tablefoottext{b}{S/N at the wavelength of the line too low.}
}
\end{table*}

\section{Nucleosynthesis and mixing models} \label{nucleo}

To interpret the measured oxygen isotopic ratios, we have
computed a set of stellar models by means of the FUNS (FUll Network
Stellar evolution) code \citep{2006NuPhA.777..311S}. In practice,
all the stable isotopes from H to F have been explicitly included
in the H-burning nuclear network coupled to the stellar structure
equations.  The reaction rates have been taken from
\citet{2010NuPhA.841...31I} and \citet{2011RvMP...83..195A}.  The
mixing length parameter, $\alpha$=1.82, was calibrated
by means of a standard solar model computed with the same set of
nuclear reaction rates, equation of state, radiative opacities, and
composition as the models used in this study.  Initial metallicity
(Z=0.015) and He abundance (Y=0.27) are the early solar values as
derived from the same standard solar model.  The solar composition
was taken from the list provided by \citet{2009LanB...4B...44L}.
For details we refer to \citet{2007A&A...462.1051P}.  
A comparison of FUNS predictions on stellar masses and cluster
ages with the values derived by \citet{1989ApJ...347..835G} showed 
good agreement within the uncertainties, so we refrained from re-determining
the stellar masses of our sample stars.
We point out
that there is a minor inconsistency in our study in the sense that
the oxygen abundance given by \citet{2009LanB...4B...44L} is 0.07
dex lower than the value of \citet{2009MmSAI..80..643C} we used for
computing the model spectra.  Since the difference is very small,
this inconsistency has negligible consequences on our analysis.

Stellar masses between 1.8 and 5.0\,$M_{\sun}$ have been investigated.
The complete set of model parameter combinations and the corresponding
results are summarized in Table\,\ref{t:modelresults}.  The first
block (first 6 rows) refers to the reference models (R), those obtained
with the recommended values of the reaction rates, the calibrated
mixing length, and the solar composition.  Then, to quantify the
theoretical uncertainties, additional models have been computed by
varying nucleosynthesis and mixing inputs, as well as the initial
abundances of the three stable O isotopes.

As usual, all the models presented here include a treatment of the
convective mixing.  Therefore, red giant models show the composition
modified by the first dredge-up.  No extra mixing induced, for example, by
rotation, thermohaline circulation, magnetic buoyancy, or gravity
wave has been considered. As is well known these processes are
hampered in red giant stars by the sharp molecular weight gradient
left by the first dredge-up. Only in stars with M $\leq 2$ $M_\odot$
does the shell H-burning reach the H discontinuity during the RGB
phase, thus smoothing down the $\mu-gradient$.  With the
possible exception of the two NGC 7789 stars, all the other stars
in our sample have masses higher than 2 $M_\odot$. In the model
with M$=1.8$ $M_\odot$, the H-burning attains the H discontinuity
when $\log(L/L_\odot)\sim 2$. The two giant stars observed in NGC
7789 are slightly brighter than this threshold\footnote{ The
luminosity may be estimated by means of $T_{\rm eff}$ and $\log g$
listed in  Table\,\ref{t:sample}.} so some extramixing has possibly
modified their compositions.  However, the observed $^{12}$C/$^{13}$C
(Table\,\ref{t:sample}) is only slightly smaller than the value
expected after the FDU and, in any case, within the error bar. The O isotopic ratios are not affected by moderate extramixing
\citep[see, e.g.,][]{2012A&A...548A..55A}.

\begin{table*}
\caption{Effects of theoretical uncertainties on $^{12}$C/$^{13}$C, $^{16}$O/$^{17}$O, and $^{16}$O/$^{18}$O.} 
\label{t:modelresults} 
\centering 
\begin{tabular}{c c c c c c c c c c} 
\hline\hline 
 & \multicolumn{4}{c}{Initial isot. ratios} & & & \multicolumn{3}{c}{Post-FDU isot. ratios}\\
label & M & $^{16}$O/$^{17}$O & $^{16}$O/$^{18}$O & $^{17}$O+p & $^{18}$O+p & $\alpha_{\rm ml}$ & $^{12}$C/$^{13}$C\tablefootmark{a} & $^{16}$O/$^{17}$O & $^{16}$O/$^{18}$O\\ 
 &  $ (M_{\sun}$) & & & (rates) & (rates) & & & & \\
\hline
\multicolumn{10}{c}{Reference models\tablefootmark{b}} \bigstrut \\
\hline
R   & 1.8 & 2696 & 499 & recom. & recom. & 1.82 & 24.5 & 440 & 684 \\
R   & 2.0 & 2696 & 499 & recom. & recom. & 1.82 & 24.2 & 262 & 703 \\
R   & 2.5 & 2696 & 499 & recom. & recom. & 1.82 & 23.8 & 208 & 711 \\
R   & 3.0 & 2696 & 499 & recom. & recom. & 1.82 & 23.5 & 253 & 709 \\
R   & 4.0 & 2696 & 499 & recom. & recom. & 1.82 & 23.2 & 338 & 703 \\
R   & 5.0 & 2696 & 499 & recom. & recom. & 1.82 & 22.9 & 381 & 704 \\
\hline              
\multicolumn{10}{c}{Nuclear reactions changed\tablefootmark{c}} \bigstrut\\
\hline
O17L   & 4.0 & 2696 & 499 & low & recom.  & 1.82 & 23.2 & 320 & 704 \\
O17H   & 4.0 & 2696 & 499 & high & recom. & 1.82 & 23.2 & 355 & 703 \\
O17L   & 2.5 & 2696 & 499 & low & recom.  & 1.82 & 23.8 & 193 & 711 \\
O17H   & 2.5 & 2696 & 499 & high & recom. & 1.82 & 23.8 & 224 & 711 \\
O18L   & 4.0 & 2696 & 499 & recom. & low  & 1.82 & 23.2 & 336 & 681 \\
O18H   & 4.0 & 2696 & 499 & recom. & high & 1.82 & 23.2 & 337 & 722 \\
O18L   & 2.5 & 2696 & 499 & recom. & low  & 1.82 & 23.8 & 210 & 693 \\
O18H   & 2.5 & 2696 & 499 & recom. & high & 1.82 & 23.8 & 210 & 727 \\
BL     & 2.5 & 2696 & 499 & hi($\gamma$),lo($\alpha$) & low     & 1.82 & 23.8 & 198 & 693 \\
\hline
\multicolumn{10}{c}{Initial composition changed\tablefootmark{d}}\bigstrut\\
\hline
C17OH     & 2.5 & 2012 & 499 & recom. & recom. & 1.82 & 23.8 & 203 & 711 \\ 
C18OH     & 2.5 & 2696 & 332 & recom. & recom. & 1.82 & 23.8 & 208 & 474 \\ 
C18OHH    & 2.5 & 2696 & 249 & recom. & recom. & 1.82 & 23.8 & 209 & 356 \\ 
C18OHH    & 3.0 & 2696 & 249 & recom. & recom. & 1.82 & 23.5 & 253 & 355 \\ 
C18OHH    & 4.0 & 2696 & 249 & recom. & recom. & 1.82 & 23.2 & 336 & 352 \\ 
C16OLL    & 2.5 & 1348 & 249 & recom. & recom. & 1 82 & 23.7 & 169 & 362 \\ 
C16OLL    & 3.0 & 1348 & 249 & recom. & recom. & 1.82 & 23.5 & 203 & 361 \\ 
C16OLL    & 4.0 & 1348 & 249 & recom. & recom. & 1.82 & 23.1 & 258 & 359 \\ 
C16OL     & 1.8 & 1617 & 299 & recom. & recom. & 1.82 & 24.5 & 343 & 413 \\
C16OL     & 2.0 & 1617 & 299 & recom. & recom. & 1.82 & 24.1 & 216 & 426 \\
C16OL     & 2.5 & 1617 & 299 & recom. & recom. & 1.82 & 23.8 & 180 & 432 \\
C16OL     & 3.0 & 1617 & 299 & recom. & recom. & 1.82 & 23.5 & 217 & 431 \\
C16OL     & 4.0 & 1617 & 299 & recom. & recom. & 1.82 & 23.1 & 278 & 428 \\
C16OL     & 5.0 & 1617 & 299 & recom. & recom. & 1.82 & 22.9 & 307 & 429 \\
\hline
\multicolumn{10}{c}{Mixing length parameters $\alpha_{\rm ml}$ changed}\bigstrut\\
\hline
 AL  & 2.5 & 2696 & 499 & recom. & recom. & 1.50 & 23.8 & 208 & 711 \\
 AH  & 2.5 & 2696 & 499 & recom. & recom. & 2.00 & 23.8 & 208 & 711 \\
\hline
\multicolumn{10}{c}{Metallicity change: Z=0.02}\bigstrut\\
\hline
 Zvar  & 2.5 & 2696 & 499 & recom. & recom. & 1.82 & 23.6 & 189 & 723 \\
\hline
\multicolumn{10}{c}{He change: Y=0.32}\bigstrut\\
\hline
 Yvar  & 2.5 & 2696 & 499 & recom. & recom. & 1.82 & 23.8 & 232 & 709\\
\hline
\end{tabular}
\tablefoot{
\tablefoottext{a}{Initial $^{12}$C/$^{13}$C always set to 89.}
\tablefoottext{b}{Reference models as obtained by assuming solar composition \citep{2009LanB...4B...44L}, 
O+p reaction rates from \citet{2010NuPhA.841...31I} and solar-calibrated mixing length.}
\tablefoottext{c}{Models obtained using the lower and upper proton capture rates of $^{17}$O or $^{18}$O
suggested by \citet{2010NuPhA.841...31I}. All other parameters equal to reference models. 
Model BL: capture rates setting chosen to minimize $^{18}$O depletion.}
\tablefoottext{d}{Initial composition changed: $^{17}$O$\times 1.34$ (C17OH), $^{18}$O$\times 1.5$ (C18OH),
$^{18}$O$\times 2$ (C18OHH), $^{16}$O$\times 0.5$ (C16OLL), $^{16}$O$\times 0.4$ (C16OL).}
}
\end{table*}

\section{Discussion}

\subsection{$^{17}$O}

All of the six sample K-giants with determined oxygen isotopic
ratios show a very similar $^{16}$O/$^{17}$O ratio between 250 and
300. Reference models, i.e. those obtained starting from a solar
composition, account for the observed values within the observational errors.
On the other hand, this agreement
does not necessarily imply that the protostellar $^{17}$O abundances
were nearly solar. In the H-burning shell, the $^{17}$O is directly
linked to the $^{16}$O through the NO cycle (see Figure \ref{f:cno}).
Therefore, the final $^{16}$O/$^{17}$O is basically fixed by the
initial abundance of the most abundant isotope of the cycle, $^{16}$O,
with variations of the initial $^{17}$O having a negligible effect.
This is demonstrated by model C17OH with an initial $^{17}$O abundance
34\% higher than the reference model but nonetheless a final
$^{16}$O/$^{17}$O ratio that is practically identical to the reference
model.  However, a variation in the initial $^{16}$O will modify
the isotope ratios after the first dredge-up.  Nevertheless, a partial
redistribution of the $^{16}$O excess or deficiency over all the
isotopes of the NO cycle mitigates the change in the final
$^{16}$O/$^{17}$O. All the six giants in our sample show clear
signatures of subsolar [$^{16}$O/H], with an average value $-0.25\pm
0.03$ (see Table \ref{t:sample}). Models C16OL demonstrate that such
a $40\%$ reduction of the initial $^{16}$O would imply a $\sim 14\%$
reduction of the final $^{16}$O/$^{17}$O ratios, which is still compatible
with the observed ratios (within $1\sigma$; see also the black-solid
curve in Figure \ref{f:17omass}).

Variations in the nuclear reaction rates have weak effects compared
to the observational uncertainties. A change of Z from 0.15 to 0.02
implies a $10\%$ increase in the $^{16}$O/$^{17}$O. On the contrary, a $10\%$ reduction is obtained if 
Y is changed from 0.27 to 0.32. 

As a whole, at variance with model predictions, the measured  $^{16}$O/$^{17}$O ratios do not show any clear evidence of a 
variation with the stellar mass. However, the small number of observations, also affected by a rather large error, hampers
 any more thorough investigation of this issue.
As outlined
in \citet{2012A&A...547A.108L}, reducing the errors is hindered by
fundamental problems in determining stellar parameters for
cool giants.  Effects of changes in the mixing length parameter are
similarly far below the observational detectability. 

\subsection{$^{18}$O}

The models show that the $^{16}$O/$^{18}$O ratio after the first
dredge-up is almost constant (for M$\ge 2$\,M$_{\sun}$). However, the five
K-giants analysed here have typical values near 350, which is about
half the value predicted by the reference models (R).  None of the
sample stars has an $^{18}$O abundance in the literature. While our
values seem to be amongst the lowest $^{16}$O/$^{18}$O ratios measured
in evolved giants, $^{16}$O/$^{18}$O values around 400 have been
measured in field giants.  \citet{1988ApJ...325..768H} report similar
values for the bright stars $\alpha$\,Ari, $\alpha$\,Ser, $\beta$\,And,
$\beta$\,Peg, and $\beta$\,UMi. The \citet{1988ApJ...325..768H}
study used different molecular lines than we have selected, which indicate
that such low ratios are probably not the result of an inappropriate
selection of spectral lines.  All but one red giant studied by
\citet{1988ApJ...325..768H} were found to have $^{16}$O/$^{18}$O
ratios $\le$\,600.

\begin{figure}
\resizebox{\hsize}{!}{\includegraphics{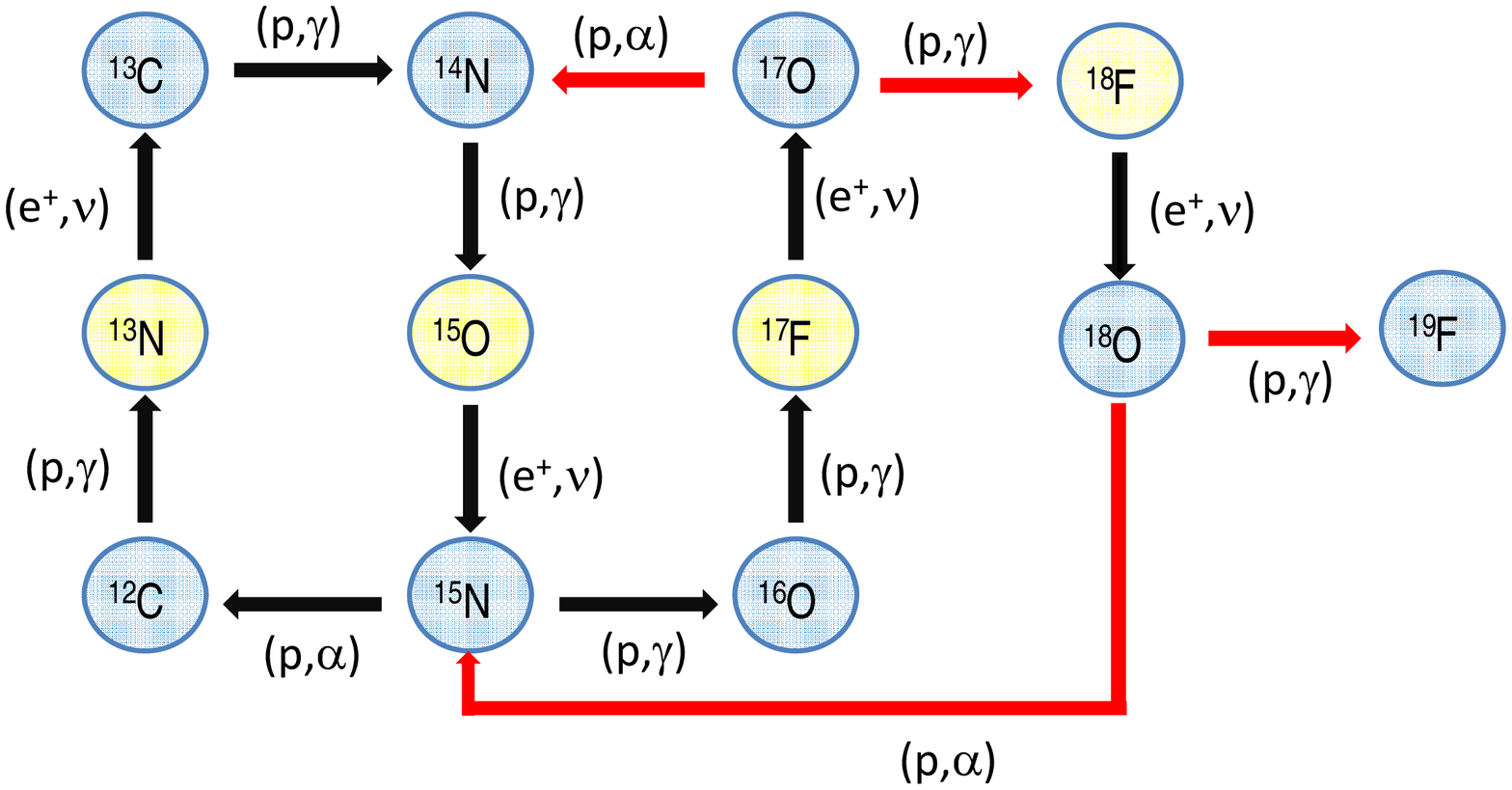}}
\caption{Illustration of the network of nuclear reactions within the CNO cycle.}
\label{f:cno}
\end{figure}

\begin{figure}
\resizebox{\hsize}{!}{\includegraphics[clip]{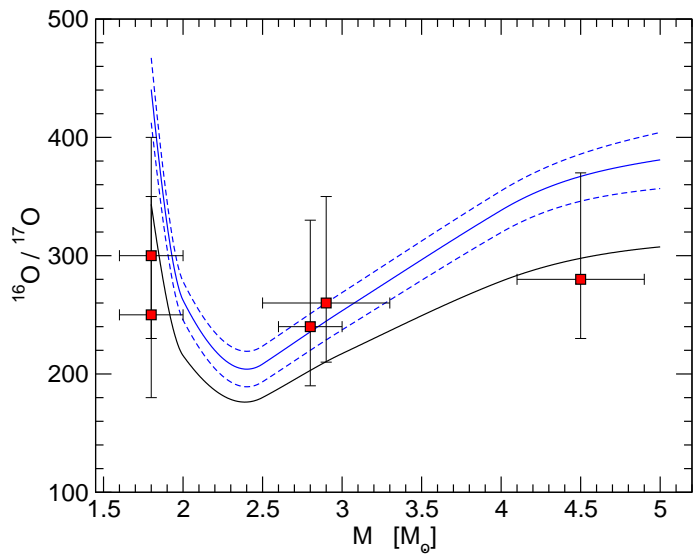}}
\caption{$^{16}$O/$^{17}$O ratios after the first dredge-up versus
stellar mass in M$_{\sun}$. Lines represent the theoretical predictions,
and the red squares are observations. 
Blue solid line: reference model (R). Blue dashed lines:
$^{17}$O proton capture rate modified within the
suggested upper and lower rates (O17L and O17H). Solid black line:
theoretical predictions as obtained by reducing the initial $^{16}$O to 
[$^{16}$O/H]=-0.22 (C16OL).}
\label{f:17omass}
\end{figure}

\begin{figure}
\resizebox{\hsize}{!}{\includegraphics[clip]{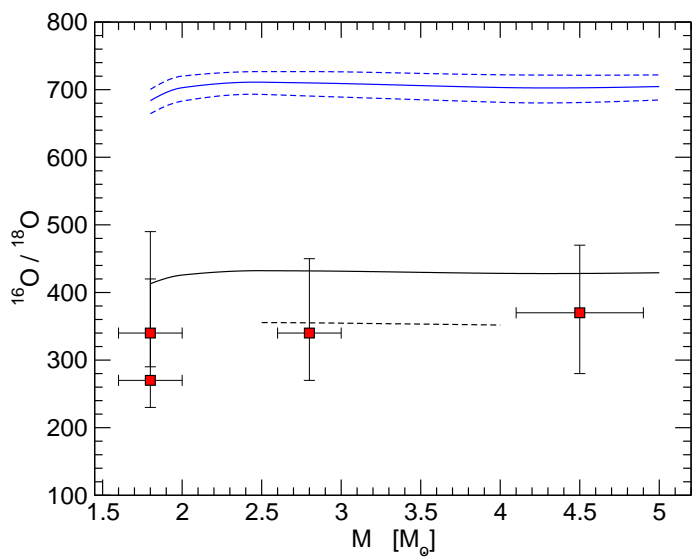}}
\caption{$^{16}$O/$^{18}$O ratios after the first dredge-up versus
stellar mass in M$_{\sun}$. Lines represent the theoretical
predictions, and the red squares are observations.
Solid blue line: reference model (R). Dashed blue lines:
$^{18}$O proton capture rate modified within the suggested upper
and lower rates (O18L and O18H). Solid black line: model predictions 
obtained by reducing the initial $^{16}$O to 
[$^{16}$O/H]=-0.22 (C16OL). Dashed black line: model with
the initial $^{18}$O abundance increased by a factor of 2 (C18OHH). 
}
\label{f:18omass}
\end{figure}

As for $^{17}$O, our tests of the uncertainties of the nucleosynthesis
model show that the only route to significantly modifying the
$^{16}$O/$^{18}$O ratio after the first dredge-up is to change the
initial abundances of individual isotopes.  At variance with the
$^{17}$O, for which only a variation in the initial $^{16}$O abundance
affects the final $^{16}$O/$^{17}$O, a modification of the initial
$^{18}$O could play an important role in interpreting the measured
$^{16}$O/$^{18}$O ratio.  Owing to the weakness of the
$^{17}$O$(p,\gamma)^{18}$F reaction, $^{18}$O remains decoupled from
the CNO cycle (see Figure \ref{f:cno}). On the other hand, it can
be easily destroyed by fast proton captures (mainly by the
$^{18}$O$(p,\alpha)^{15}$N reaction) so that the $^{16}$O/$^{18}$O
ratio should increase after the first dredge-up\footnote {As
previously noted $^{16}$O is practically unaffected by the first
dredge-up.}. Our models show that for any choice of the initial
$^{16}$O,$^{18}$O abundance pair and for any initial mass, the final
$^{16}$O/$^{18}$O will always be $\sim 1.4$ times greater than the
corresponding initial value.  For this reason, starting from a solar
value of the $^{16}$O/$^{18}$O ratio, $\sim 500$, it is impossible
to account for the measured values.  Our models require that either
the $^{16}$O or the $^{18}$O be varied in order to match the
observations. For instance, model C18OHH with an initial $^{18}$O twice
solar abundance matches the observed $^{16}$O/$^{18}$O.  However,
this is also the result when using model C16OLL with an initial $^{16}$O
abundance half solar.  It follows that combinations of these two
assumptions also provide a good reproduction of the observations.
For instance, assuming a $40\%$ reduction of the initial $^{16}$O,
as implied by the observed average [$^{16}$O/H], results in
$^{16}$O/$^{18}$O, which is in reasonable agreement with the measured value
(within $1\sigma$; Figure \ref{f:18omass}).
The agreement can be improved further by increasing the initial
$^{18}$O about 10\%.

\subsection{Implications of the results}      

By combining [$^{16}$O/H], $^{16}$O/$^{17}$O, and $^{16}$O/$^{18}$O
measurements in six giants belonging to galactic OC of
known ages, we can infer the O isotopic composition of the respective
protostellar nebulae. In particular, the optimal protostellar
abundances modelled from the observed stellar abundances imply
subsolar $^{16}$O, i.e. $-40\%$ on the average, and slightly
supersolar $^{18}$O, i.e. $+10\%$.  The $^{16}$O/$^{18}$O ratios
in the protostellar gas are reduced by a factor of $\sim 1.4$ from
the values measured in the red giant stars.  No direct information
on the protostellar $^{17}$O can be obtained from our observations.

Is this picture that arises from our measurements in OC giants compatible
with the predictions of extant models of galactic chemical 
evolution\footnote{Isotopic ratios of oxygen measured in stars are typically
given as $^{16}$O/$^{xx}$O in the literature. Accordingly, we used this
format of the isotopic
ratio in the first part of the paper. Galactic chemical evolution models,
however, often give the reversed ratio $^{xx}$O/$^{16}$O. Therefore, we
decided to use $^{xx}$O/$^{16}$O in this section to allow for an easier comparison
with the literature.}
(GCE)?  $^{16}$O is a primary product of the He burning. It is
produced through the main He-burning chain: $3\alpha \rightarrow
^{12}$C$+\alpha \rightarrow ^{16}$O.  $^{18}$O is also produced in
He-burning regions as a secondary product synthesized through the
following chain: $^{14}$N($\alpha$,$\gamma$)$^{18}$F($\beta$)$^{18}$O,
where the abundance of $^{14}$N is equivalent, in practice, to the
original amount of C+N+O.  The most efficient polluters of both
$^{16}$O and $^{18}$O are massive stars exploding as core-collapse
supernovae.  $^{17}$O is a secondary product of the H burning (CNO
cycle, see Figure \ref{f:cno}).  Therefore, in addition to massive
stars, the winds of intermediate and low mass stars contribute on a
longer timescale to the oxygen pollution of the ISM by contributing
$^{17}$O.  In this framework, it is not a surprise that GCE models generally predict an increase with time in both
the $^{16}$O/$^{18}$O and the $^{16}$O/$^{17}$O, as expected for the ratio of a primary to a secondary product. 
Less straightforward is the evolution of the ratio of the 
two secondary O isotopes. In general, when two secondary isotopes are produced by the same class of stars, their
ratio should remain constant. However, this is not the case for $^{18}$O/$^{17}$O,
which is expected to decrease when low and intermediate mass stars begin to contribute to the 
pollution of the lightest isotope \citep{1988ApJ...334..191C,1996A&A...309..760P,2008Meyer,
2011MNRAS.414.3231K,2012M&PS...47.2031N}.

\begin{figure}
\resizebox{\hsize}{!}{\includegraphics[clip]{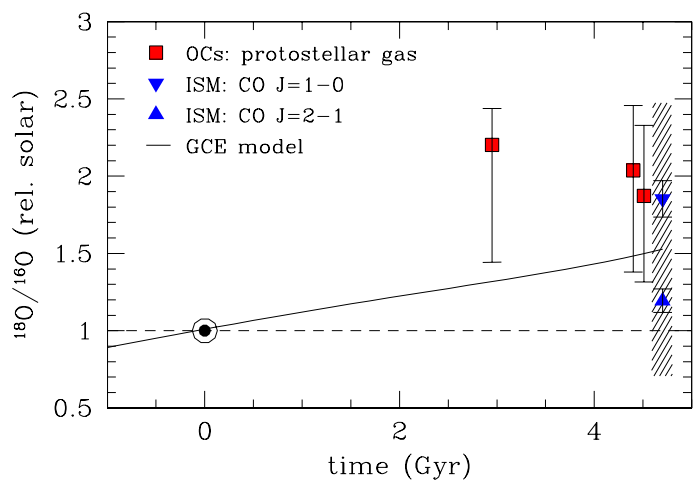}}
\caption{Evolution of $^{18}$O/$^{16}$O with time. The isotopic ratio 
is plotted relative to the solar value. The zero point of the time
axis is defined by the formation of the sun. The solid black line
indicates the predictions from galactic chemical evolution (GCE) models.
Red boxes mark our estimates of $^{18}$O/$^{16}$O in the parent nebulae of the 
three OCs NGC 7789, NGC 1545, and Tr2. 
The two measurements for NGC 7789 have been combined to one point.
Blue symbols indicate ISM values.
The shaded area is representative of the whole range of ISM measurements.
}
\label{f:gce}
\end{figure}

In Figure \ref{f:gce} predictions of the \citet{1996A&A...309..760P}
GCE model are compared with our estimates of $^{18}$O/$^{16}$O in
the parent nebulae of the OC.  The solid line represents
the predicted evolution of the $^{18}$O/$^{16}$O ratio over the
past 6 Gyr.  The CGE model refers to abundances in the interstellar
medium (ISM) at the current solar galactocentric radius,
$R_{GC}=8.5$\,kpc.  The $^{18}$O/$^{16}$O ratio is plotted relative
to the solar value while time is scaled to the epoch of the solar
system formation.  Red squares show our isotope ratio estimates for
the three clusters NGC 7789, NGC 1545, and Tr 2.  In spite of the
large uncertainties, it appears that the primordial gas of the three
clusters underwent a faster increase in the $^{18}$O abundance than
predicted from the GCE model.  This figure does not change by
using the more recent \citet{2011MNRAS.414.3231K} GCE model.  The
galactocentric distance of the three clusters is only slightly
greater than solar, between 9\,kpc and 9.7\,kpc. Although migration of
the Sun and the clusters\footnote {Effects of migration are probably
small for the two younger clusters, NGC 1545 and Tr 2, while it
cannot be excluded in the case of NGC 7789 (1.7 Gyr old).} from
small galactocentric radius cannot be excluded, the birthplaces
should be within $6<R_{GC}<10$\,kpc.

Oxygen isotopic ratios in molecular clouds located in this portion
of the Galaxy are available from infrared and radio observations
\citep{2008A&A...487..237W,2005ApJ...634.1126M, 2005A&A...437..957P}.
\citet{2012M&PS...47.2031N} have derived $^{18}$O/$^{16}$O for
molecular clouds by combining $^{13}$C$^{16}$O/$^{12}$C$^{18}$O
reported by \citet{2008A&A...487..237W} with the galactic
$^{12}$C/$^{13}$C gradient obtained by \citet{2005ApJ...634.1126M}
from CO observations.  The two triangles in Figure \ref{f:gce}
represent the weighted averages of these measurements, as obtained
from two different molecular lines, CO J=1-0 and CO J=2-1.  Only
molecular clouds for $6<R_{GC}<10$\,kpc have been considered.  Other
measurements, derived from OH lines \citep{2005A&A...437..957P} and
formaldehyde data \citep{1994ARA&A..32..191W}, are within the range
covered by the CO measurements.  In the same figure, the shaded
area is representative of the whole set of available measurements
for the ISM lying on the galactic disk approximately between 6 and
10\,kpc \citep[see also figure 2 in][]{2012M&PS...47.2031N}.  Compared
to the ISM, the $^{18}$O/$^{16}$O ratios we have derived for three
OCs are close to the upper bound of the available measurements.

Although we cannot constrain the protostellar $^{17}$O from OC
giant observations, molecular clouds located between 6 and 1\,kpc
from the galactic centre show that the $^{18}$O/$^{17}$O ratio
ranges between 3 and 5, with an average value $4.16\pm 0.09$
(\citet{2008A&A...487..237W}; see also \citet{1981ApJ...249..518P}
and \citet{2005ApJ...634.1126M}).  
Since the ratio
of these two secondary isotopes can either remain constant or decrease with time
and since our measurements indicate slightly supersolar $^{18}$O
in the ISM from which the host clusters were born, a similar or
even larger enhancement of $^{17}$O is expected, otherwise the
initial $^{18}$O/$^{17}$O would be definitely too large compared
to the typical values found in the galactic disk.

\section{Conclusions}

We measured the ratios of the three main isotopes of oxygen in the
atmospheres of six evolved K giants. Except for one the stars are
members of stellar clusters and therefore have well-defined masses.
The $^{16}$O/$^{17}$O and $^{16}$O/$^{18}$O ratios are compared
with model predictions.  There is agreement between observations
and predictions for the $^{16}$O/$^{17}$O ratio.  The dependency
of $^{16}$O/$^{17}$O on mass, expected from nucleosynthesis models,
is not observed. However, the predicted variation
is comparable to the data errors.
The observed $^{16}$O/$^{18}$O ratio can be brought
into agreement with model predictions if the initial isotopic ratio
is about half of the solar value.  Uncertainties in proton capture
rates and mixing length cannot account for the observed difference.
By combining this result with our estimation of the [$^{16}$O/H],
we conclude that a moderate enhancement of the initial $^{18}$O
abundance relative to the solar value in combination with a subsolar
$^{16}$O abundance can provide a good match to the observations.
Finally, to maintain the protostellar $^{18}$O/$^{17}$O in agreement
with the values typically measured in nearby molecular clouds, we
also infer a moderate enhancement of the initial $^{17}$O abundance.

\begin{acknowledgements}
We are indebted to Larry Nittler for providing a computer-readable
table with his derivation of the $^{16}$O/$^{18}$O in molecular
clouds. We thank the anonymous referee for the very valuable 
comments T.L and W.N. were supported by Austrian Science Fund FWF
under project number P23737-N16. W.N. was also supported by project
P21988-N16.  B.A. acknowledges the support from the {\em project
STARKEY} funded by the ERC Consolidator Grant, G.A. n.~615604.  This
paper is based on observations obtained at the Kitt Peak National
Observatory (NOAO Prop.\,ID:2013B-0218; PI: T.\,Lebzelter).  The
Kitt Peak National Observatory is part of the National Optical
Astronomy Observatory, which is operated by the Association of
Universities for Research in Astronomy (AURA) under cooperative
agreement with the National Science Foundation.  
\end{acknowledgements}

\bibliographystyle{aa} 
\bibliography{oisotop}

\begin{appendix}
\section{Line data}

\begin{table*}
\caption{Molecular lines of $^{12}$C$^{16}$O (selection), $^{12}$C$^{17}$O, $^{12}$C$^{18}$O, 
$^{13}$C$^{16}$O, and $^{16}$OH used in the analysis.} 
\label{t:linelist} 
\centering 
\begin{tabular}{l c r r r} 
\hline\hline 
Molecule\tablefootmark{a} & Transition & wavelength\tablefootmark{b} & E\tablefootmark{c} & $gf$\\
 & & [{\AA}] & [cm$^{-1}$] & \\
\hline\bigstrut
$^{12}$C$^{16}$O & 2-0 R1 & 23432.693 & 3.845 & 1.802E-07\\
 & 4-2 R41 & 23540.840 & 7492.257 & 2.962E-05\\
 & 4-2 R60 & 23547.324 & 11086.589 & 4.871E-05\\
 & 4-2 R38 & 23553.885 & 7043.948 & 2.698E-05\\
 & 4-2 R24 & 23664.988 & 5390.375 & 1.583E-05\\
 & 3-1 R4 & 23667.714 & 2181.369 & 1.392E-06\\
 & 2-0 P9 & 23680.105 & 172.978 & 7.605E-07\\
\bigstrut
$^{12}$C$^{17}$O  & 2-0 R14 & 23502.671 & 393.276 & 1.407E-06\\
 & 2-0 R30 & 23316.224 & 1737.760 & 3.223E-06\\
 & 2-0 R29 & 23324.897 & 1625.950 & 3.099E-06\\
 & 2-0 R28 & 23333.966 & 1517.827 & 2.976E-06\\
 & 2-0 R26 & 23353.313 & 1312.658 & 2.735E-06\\
\bigstrut $^{12}$C$^{18}$O & 2-0 R23 & 23656.947 & 1009.011 & 2.298E-06\\
$^{13}$C$^{16}$O & 2-0 R68 & 23501.045 & 8501.143 & 8.980E-06\\
 & 2-0 R69 & 23507.818 & 8747.480 & 9.172E-06\\
 & 2-0 R70 & 23515.015 & 8997.172 & 9.367E-06\\
 & 2-0 R71 & 23522.680 & 9250.210 & 9.565E-06\\
 & 2-0 R76 & 23567.025 & 10565.262 & 1.059E-05\\
 & 2-0 R19 & 23660.695 & 697.622 & 1.878E-06\\
\bigstrut
$^{16}$OH & 4-2 P$_{1f}$ 6.5 & 16477.325 & 7683.8196 & 1.800E-05\\
 & 2-0 P$_{1f}$ 16.5 & 16460.532 & 4915.1296 & 1.114E-05\\
 & 2-0 P$_{2e}$ 15.5 & 16452.548 & 4939.7718 & 1.047E-05\\
\hline
\end{tabular}
\tablefoot{
\tablefoottext{a}{CO line data from \citet{1994ApJS...95..535G}; OH lines from 
\citet{2009JQSRT.110..533R}}
\tablefoottext{b}{vacuum}
\tablefoottext{c}{lower state term energy}
}
\end{table*}

\end{appendix}

\end{document}